\def\year{2019}\relax
\documentclass[letterpaper]{article} 
\usepackage{aaai19}  
\usepackage{times}  
\usepackage{helvet}  
\usepackage{courier}  
\usepackage{url}  
\usepackage{graphicx}  
\usepackage{float}
\usepackage{amsmath, amsthm, amssymb, amsfonts}
\usepackage{centernot}

\begin{document}
%
\title{Learning to Address Health Inequality in the United States with a Bayesian Decision Network}

\author{Tavpritesh Sethi\textsuperscript{1,2}, 
Anant Mittal\textsuperscript{2}, 
Shubham Maheshwari\textsuperscript{2}, 
Samarth Chugh\textsuperscript{3}\\
\textsuperscript{1}Stanford University, USA\\
\textsuperscript{2}Indraprastha Institute of Information Technology, Delhi, India\\
\textsuperscript{3}Netaji Subhas University of Technology, Delhi, India\\
tavsethi@stanford.edu, tavpriteshsethi@iiitd.ac.in
}

\maketitle
\begin{abstract}
\begin{quote}
Life-expectancy is a complex outcome driven by genetic, socio-demographic, environmental and geographic factors. Increasing socio-economic and health disparities in the United States are propagating the longevity-gap, making it a cause for concern. Earlier studies have probed individual factors but an integrated picture to reveal quantifiable actions has been missing. Amidst growing concerns about the further widening of healthcare inequality and differential access created by Artificial Intelligence, it is imperative to explore its potential for illuminating biases and enabling transparent policy decisions. In this work, we reveal actionable interventions for decreasing the longevity-gap in the United States by analyzing a County-level data resource with healthcare, socio-economic, behavioral, education and demographic features. We learn an ensemble-averaged structure, draw inferences using the joint probability distribution and extend it to a Bayesian Decision Network for identifying policy actions. We draw quantitative estimates for the positive roles of diversity, preventive-care quality and stable-families within the unified framework of our decision network. Finally, we make this analysis and dashboard available as an interactive web-application for enabling users and policy-makers to validate our insights on bridging the longevity-gap and explore the ones beyond reported in this work.
\end{quote}
\end{abstract}

\section{Introduction}
\noindent Inequities in healthcare impose an estimated burden of \$300 billion per year in the United States \cite{LaVeist2011}. Longevity-gap is a collective effect of inequities such as economic, racial, ethnic, gender and environmental that lead to a decrease in life-expectancy. While the effects of inequities on longevity have been a subject of study for social scientists for decades, the society is now presented with an opportunity (or a threat) for the use of Artificial Intelligence (AI) in addressing (or propagating) healthcare inequities. Most of the AI research in healthcare has focused on developing discriminative models for automated diagnosis and predictions. Despite many promising applications, and the raised bar for demonstrating value in Clinical Trials \cite{Abramoff2018}, the opaque nature of many of these models has raised concerns about propagation of blind-spots resulting from human-biases \cite{Cabitza2017,AS2018,Keane2018}. In this study we argue for the use of expressive, interpretable and explainable AI for understanding healthcare inequities themselves and for learning actions that may help mitigate these. We learn the complex interplay of factors that influence life-expectancy through a data-driven Bayesian Decision Network, i.e. a joint probabilistic graphical model (PGM) learned from data and extended to a decision framework for revealing policy actions. The key motivation for this study was to create a unified model that integrates geographical, socio-economic, behavioral, demographic and healthcare indices at county-level resolution in order to discover the skeleton and key drivers of the longevity-gap. Since graphical models are intuitive and interpretable, we enhance the utility of our model with an interactive web-application available to users for exploring and discovering further insights from our model. 
\section{Determinants of the Longevity-gap}
\noindent While access to medical care is the most tangible factor for improving longevity, the latter is determined by a more nuanced interplay of social, behavioral, and economic factors. This interplay has been difficult to de-convolve so far because the non-availability of integrated datasets gathered over a long time-span and a high spatial resolution. Hence, various theories and models have assessed these factors in isolation. \textit{The Health Inequity Project}, \cite{Chetty2016} compiled and released granular data at the County, Commute Zone, Core Based Statistical Area and State levels, and demonstrated the longevity gap attributable to income disparity in the United States. Income data from 1.4 billion Tax Records were combined with County-level Mortality (Census), Healthcare (Dartmouth Atlas, Small Area Insurance Estimates) Health-behaviors (CDC Behavioral Risk Factor Surveillance System, BRFSS), and Education (National Center for Education Statistics Common Core of Data \& Integrated Postsecondary Education Data System) for the period covering 1999-2014, thus generating the most fine-grained data resource on socio-demographic determinants of longevity so far. Standard statistical analysis revealed significant correlations with longevity, however a bigger picture integrating the heterogeneous data into a single model has been lacking. Next, we review the formulation of problem as Bayesian structure-learning, inference, and decision networks, which we used in this study as a unified model to draw estimates and policy for addressing the longevity-gap. The problem was set up as a data-driven Bayesian decision network learning for discrete policy decisions and inferences.

\section{Bayesian Decision Networks}
We first establish the notational conventions before giving the definitions of Bayesian Networks (BNs) and Bayesian Decision Networks (BDNs). We will use $v$ to represent a node and $V$ to represent a set of nodes. The parent nodes of $v$ are represented as $pa(v)$.
\subsection{Bayesian Network}

A Bayesian Network \cite{Pearl1985a}  is a triple, $N = \{X, G, P\}$  defined over a set of random variables, $X$, consisting of a directed acyclic graph (DAG), $G$, together with the conditional probability distribution $P$. The DAG, $G = (V, E)$ is made up of vertices $V = \{v_1, v_2, ..., v_n\}$ and directed edges $E \sqsubseteq V \times V$ specifying the assumptions of conditional independence between random variables according to the d-separation criterion \cite{Geiger1990}. The set of conditional probability distributions, $P$, contains $P(x_v | x_{pa(v)})$ for each random variable $x_v\in X$. Since a Bayesian Network encodes the conditional independencies based on the d-separation criterion, it provides a compact representation for the data. The joint-probability distribution over a set of variables, can be factored as the product of probabilities of each node $v$ conditioned upon its parents. 

\begin{equation} 
\begin{split}
P(X) = \prod_{v\in V}{P(x_v | x_{pa(v)})}
\end{split}
\end{equation}

\subsection{Structure-learning}
The structure of a Bayesian Network represents conditional independencies and can be specified either by an expert or can be learned from data.  In data-driven structure-learning, the goal is to identify a model representing the underlying joint distribution of the data, $P$. For simplification, we assume the \textit{faithfulness} criterion, that is, the underlying joint probability distribution $P$ can be represented as a DAG. Structure-learning of the DAG can then be carried out with one of the three classes of algorithms: constraint-based, score based or hybrid algorithms. Since constraint-based methods rely on individual tests of independence, they are known to suffer from the problem of being sensitive to individual failures \cite{Koller2009}. Score based methods view the structure learning as a model-selection problem and are implemented as search and score-based strategy. These are computationally more expensive because of the super-exponential space of models, however recent theoretical developments have made these tractable \cite{Koller2009}. In this study, hill-climbing search was used along with the \textit{Bayesian Information Criterion} (BIC) scoring function. The scoring function is a measure of \textit{goodness} of each Bayesian Network $N=\{X,G,P\}$, for representing the data, $D$. The BIC scoring function takes the form

\begin{equation}
bic(G : D) = l(\theta^G:D) - (Dim[G] (log N/2)) 
\end{equation}

where $N$ is the sample size of the data, $D$, and $l(\theta^G:D)$ is the likelihood fit to the data. We chose the BIC scoring function as it penalizes complex models over sparse ones and is a good trade-off between likelihood and the model complexity, thus reducing over-fitting.

The \textit{hill-climbing} search algorithm is a local strategy wherein each step, the next candidate neighboring structure of the current candidate is selected based on the highest score and is described as follows.
\begin{enumerate}
	\item Initialize the structure $G$.
    \item Repeat
    \begin{enumerate}
    	\item Add, remove or reverse an arc at a time for modifying $G$ to generate a set \textbf{G} of candidate structures.
        \item Compute the score of each candidate structure in \textbf{G}.
        \item Pick the change with highest score and set it as new $G$
    \end{enumerate}
    Until the score cannot be improved.
    \item Return $G$.
\end{enumerate}

\subsection{Inference}
The next task is to parametrize the learned structure by computing posterior marginal distribution of an unobserved variable $x_{v_j} \in X$ given an evidence $e=\{e_1, e_2,.. e_m\}$ on a set of variables $X(e)$. The prior marginal distribution $P(x_{v_j})$ can be computed as

\begin{equation} 
\begin{split}
P(x_{v_j}) & = \sum_{x \in X \setminus \{x_{v_j}\}} {P(X)}
\end{split}
\end{equation}
Using (1) in (3), we get
\begin{equation} \label{eq2}
\begin{split}
P(x_{v_j}) = {\sum_{x \in X \setminus \{x_{v_j}\}} \prod_{x_v \in X} P(x_v| x_{pa(v)})}
\end{split}
\end{equation}

Further evidence $e$ can be incorporated as
\begin{equation} 
\begin{split}
P(x_{v_j}|e) & = P(x_{v_j}, e) / P(e) \\
& \propto P(x_{v_j}, e) \\
& = \sum_{y \in X \setminus \{x_{v_j}\}} P(X,e) \\
& = \sum_{y \in X \setminus \{x_{v_j}\}}  \prod_{x_{v_i} \in X} P(x_{v_i}| x_{pa(v_i)}) E_e \\
& = \sum_{y \in X \setminus \{x_{v_j}\}}  \prod_{x_{v_i} \in X} P(x_{v_i}| x_{pa(v_i)}) \prod_{x \in X(e))} {E_x} \\
\end{split}
\end{equation}

For each $x_{v_j}\notin X(e)$, where $E_x$ is the evidence function for $x \in X(e)$ and $v_i$ is the node representing $x_{v_i}$. The likelihood is then defined as, 
\begin{equation}
\begin{split}
L(x_{v_j}|e) & = P(e|x_{v_j}) \\
& = \sum_{y \in X \setminus \{x_{v_j}\}}  \prod_{i \neq j} P(x_{v_i}| x_{pa(v_i)}) \prod_{x \in X(e))} {E_x} \\
\end{split}
\end{equation}
that is, the likelihood function of $x_{v_j}$ given $e$. Application of Bayes rule and using (4) and (6), then yields 

\begin{equation}
P(x_{v_j}|e) \propto L(x_{v_j}|e) P(x_{v_j}) 
\end{equation}

where the proportionality constant $P(e)$ can be computed from $P(X,e)$ by summation over $X$ and $P(x_{v_j})$ may be obtained by inference over empty set of evidence.\\

This parametrization of the learned structure with posterior marginal conditional probabilities allows to make Inference and the task is called \textit{inference-learning}. Inference learning is an extremely useful step that allow us to estimate marginal probabilities after setting evidences in the network. Depending upon the size of the network, inference-learning can be carried out with \textit{exact}, i.e. closed form or \textit{approximate} i.e. based upon Monte Carlo simulation methods. The details of these can be found in standard texts \cite{Koller2009}. In this work, we derived both exact estimates using the clique-tree algorithm and approximate estimates using \textit{rejection sampling} to estimate conditional probabilities.\vspace{\baselineskip}

\subsection{Bayesian Decision Network}
A Bayesian Network can be extended to a decision network (BDN) with the addition of \textit{utility} nodes $U$. A BDN extends a BN in the sense that a BN is a probabilistic network for belief update, where as a BDN is a probabilistic network for decision making under uncertainty. Formally defined, a BDN is a quadruple BDN:  \( N= \{ X,G,P,U \}  \) and  has three types of nodes \begin{itemize}
	\item Chance nodes ( \( X_{C} \) )\ :\ Nodes which represent events not controlled  by the decision  maker.\par

	\item Decision nodes ( \( X_{D} \) ) : Nodes which represent actions under direct control of the decision maker.\par

	\item Utility nodes (  \( X_{U}) :  \) Nodes which represent the decision maker's preference. These nodes can not be parents of chance or decision nodes.\par

\end{itemize}
\par

A decision maker interested in choosing best possible actions can either specify a structure and marginal probability distributions or use the model learned with structure and inference learning. The decision maker then ascribes a utility functions to particular states of a node. The objective of decision analysis is then to identify the decision options that maximize the expected utility. \par

The expected utility for each decision option is computed and the one which has the maximum utility output is returned. The mathematical formulation for the same can be given as follows: Let  \( A \)  be the decision variable with options  \( a_{1},a_{2}......a_{m} \) and  \( H \) is the hypothesis with states  \( h_{1},h_{2},........h_{n} \)  and  \(  \varepsilon  \) is a set of observations in the form of evidence, then the utility of an outcome (\( a_{i},h_{j}\))  is given by  \( U (a_{i},h_{j} )\) where  \( U \left( . \right)  \)  is the utility function and the expected utility is given as:\par

\vspace{\baselineskip}
\begin{equation}
EU( a_{i})  =  \sum _{j}^{}U( a_{i},h_{j}) P( h_{j} \vert  \varepsilon) _{}
\end{equation}

Where  \( P \left( . \right)  \)  is our belief in  \( H \) given  \(  \varepsilon  \) . We select the decision option  \( a^{\ast} \) which maximizes the expected utility such that:\par

\vspace{\baselineskip}
\begin{equation}
a^{\ast}= arg max_{a  \epsilon  A}EU(a)
\end{equation}

\vspace{\baselineskip}

\vspace{\baselineskip}

\section{Experiments}
\subsection{Dataset}
\def\year{2019}\relax
\documentclass[letterpaper]{article} 
\usepackage{aaai19}  
\usepackage{times}  
\usepackage{helvet}  
\usepackage{courier}  
\usepackage{url}  
\usepackage{graphicx}  
\frenchspacing  
\setlength{\pdfpagewidth}{8.5in}  
\setlength{\pdfpageheight}{11in}  
\usepackage{float}
\usepackage{amsmath, amsthm, amssymb, amsfonts}
\usepackage{centernot}


\pdfinfo{
/Title (Learning to Address Health Inequality in the United States with a Bayesian Decision Network)
/Author (Tavpritesh Sethi, Anant Mittal, Shubham Maheshwari, Samarth Chugh)}
\setcounter{secnumdepth}{0}  
 \begin{document}
%
\title{Learning to Address Health Inequality in the United States with a Bayesian Decision Network}

\author{Tavpritesh Sethi\textsuperscript{1,2}, 
Anant Mittal\textsuperscript{2}, 
Shubham Maheshwari\textsuperscript{2}, 
Samarth Chugh\textsuperscript{3}\\
\textsuperscript{1}Stanford University, USA\\
\textsuperscript{2}Indraprastha Institute of Information Technology, Delhi, India\\
\textsuperscript{3}Netaji Subhas University of Technology, Delhi, India\\
tavsethi@stanford.edu, tavpriteshsethi@iiitd.ac.in
}

\maketitle
\begin{abstract}
\begin{quote}
Life-expectancy is a complex outcome driven by genetic, socio-demographic, environmental and geographic factors. Increasing socio-economic and health disparities in the United States are propagating the longevity-gap, making it a cause for concern. Earlier studies have probed individual factors but an integrated picture to reveal quantifiable actions has been missing. Amidst growing concerns about the further widening of healthcare inequality and differential access created by Artificial Intelligence, it is imperative to explore its potential for illuminating biases and enabling transparent policy decisions. In this work, we reveal actionable interventions for decreasing the longevity-gap in the United States by analyzing a County-level data resource with healthcare, socio-economic, behavioral, education and demographic features. We learn an ensemble-averaged structure, draw inferences using the joint probability distribution and extend it to a Bayesian Decision Network for identifying policy actions. We draw quantitative estimates for the positive roles of diversity, preventive-care quality and stable-families within the unified framework of our decision network. Finally, we make this analysis and dashboard available as an interactive web-application for enabling users and policy-makers to validate our insights on bridging the longevity-gap and explore the ones beyond reported in this work.
\end{quote}
\end{abstract}

\section{Introduction}
\noindent Inequities in healthcare impose an estimated burden of \$300 billion per year in the United States \cite{LaVeist2011}. Longevity-gap is a collective effect of inequities such as economic, racial, ethnic, gender and environmental that lead to a decrease in life-expectancy. While the effects of inequities on longevity have been a subject of study for social scientists for decades, the society is now presented with an opportunity (or a threat) for the use of Artificial Intelligence (AI) in addressing (or propagating) healthcare inequities. Most of the AI research in healthcare has focused on developing discriminative models for automated diagnosis and predictions. Despite many promising applications, and the raised bar for demonstrating value in Clinical Trials \cite{Abramoff2018}, the opaque nature of many of these models has raised concerns about propagation of blind-spots resulting from human-biases \cite{Cabitza2017,AS2018,Keane2018}. In this study we argue for the use of expressive, interpretable and explainable AI for understanding healthcare inequities themselves and for learning actions that may help mitigate these. We learn the complex interplay of factors that influence life-expectancy through a data-driven Bayesian Decision Network, i.e. a joint probabilistic graphical model (PGM) learned from data and extended to a decision framework for revealing policy actions. The key motivation for this study was to create a unified model that integrates geographical, socio-economic, behavioral, demographic and healthcare indices at county-level resolution in order to discover the skeleton and key drivers of the longevity-gap. Since graphical models are intuitive and interpretable, we enhance the utility of our model with an interactive web-application available to users for exploring and discovering further insights from our model. 
\section{Determinants of the Longevity-gap}
\noindent While access to medical care is the most tangible factor for improving longevity, the latter is determined by a more nuanced interplay of social, behavioral, and economic factors. This interplay has been difficult to de-convolve so far because the non-availability of integrated datasets gathered over a long time-span and a high spatial resolution. Hence, various theories and models have assessed these factors in isolation. \textit{The Health Inequity Project}, \cite{Chetty2016} compiled and released granular data at the County, Commute Zone, Core Based Statistical Area and State levels, and demonstrated the longevity gap attributable to income disparity in the United States. Income data from 1.4 billion Tax Records were combined with County-level Mortality (Census), Healthcare (Dartmouth Atlas, Small Area Insurance Estimates) Health-behaviors (CDC Behavioral Risk Factor Surveillance System, BRFSS), and Education (National Center for Education Statistics Common Core of Data \& Integrated Postsecondary Education Data System) for the period covering 1999-2014, thus generating the most fine-grained data resource on socio-demographic determinants of longevity so far. Standard statistical analysis revealed significant correlations with longevity, however a bigger picture integrating the heterogeneous data into a single model has been lacking. Next, we review the formulation of problem as Bayesian structure-learning, inference, and decision networks, which we used in this study as a unified model to draw estimates and policy for addressing the longevity-gap. The problem was set up as a data-driven Bayesian decision network learning for discrete policy decisions and inferences.

\section{Bayesian Decision Networks}
We first establish the notational conventions before giving the definitions of Bayesian Networks (BNs) and Bayesian Decision Networks (BDNs). We will use $v$ to represent a node and $V$ to represent a set of nodes. The parent nodes of $v$ are represented as $pa(v)$.
\subsection{Bayesian Network}

A Bayesian Network \cite{Pearl1985a}  is a triple, $N = \{X, G, P\}$  defined over a set of random variables, $X$, consisting of a directed acyclic graph (DAG), $G$, together with the conditional probability distribution $P$. The DAG, $G = (V, E)$ is made up of vertices $V = \{v_1, v_2, ..., v_n\}$ and directed edges $E \sqsubseteq V \times V$ specifying the assumptions of conditional independence between random variables according to the d-separation criterion \cite{Geiger1990}. The set of conditional probability distributions, $P$, contains $P(x_v | x_{pa(v)})$ for each random variable $x_v\in X$. Since a Bayesian Network encodes the conditional independencies based on the d-separation criterion, it provides a compact representation for the data. The joint-probability distribution over a set of variables, can be factored as the product of probabilities of each node $v$ conditioned upon its parents. 

\begin{equation} \label{eq1}
\begin{split}
P(X) = \prod_{v\in V}{P(x_v | x_{pa(v)})}
\end{split}
\end{equation}

\subsection{Structure-learning}
The structure of a Bayesian Network represents conditional independencies and can be specified either by an expert or can be learned from data.  In data-driven structure-learning, the goal is to identify a model representing the underlying joint distribution of the data, $P$. For simplification, we assume the \textit{faithfulness} criterion, that is, the underlying joint probability distribution $P$ can be represented as a DAG. Structure-learning of the DAG can then be carried out with one of the three classes of algorithms: constraint-based, score based or hybrid algorithms. Since constraint-based methods rely on individual tests of independence, they are known to suffer from the problem of being sensitive to individual failures \cite{Koller2009}. Score based methods view the structure learning as a model-selection problem and are implemented as search and score-based strategy. These are computationally more expensive because of the super-exponential space of models, however recent theoretical developments have made these tractable \cite{Koller2009}. In this study, hill-climbing search was used along with the \textit{Bayesian Information Criterion} (BIC) scoring function. The scoring function is a measure of \textit{goodness} of each Bayesian Network $N=\{X,G,P\}$, for representing the data, $D$. The BIC scoring function takes the form

\begin{equation}
bic(G : D) = l(\theta^G:D) - (Dim[G] (log N/2)) 
\end{equation}

where $N$ is the sample size of the data, $D$, and $l(\theta^G:D)$ is the likelihood fit to the data. We chose the BIC scoring function as it penalizes complex models over sparse ones and is a good trade-off between likelihood and the model complexity, thus reducing over-fitting.

The \textit{hill-climbing} search algorithm is a local strategy wherein each step, the next candidate neighboring structure of the current candidate is selected based on the highest score and is described as follows.
\begin{enumerate}
	\item Initialize the structure $G$.
    \item Repeat
    \begin{enumerate}
    	\item Add, remove or reverse an arc at a time for modifying $G$ to generate a set \textbf{G} of candidate structures.
        \item Compute the score of each candidate structure in \textbf{G}.
        \item Pick the change with highest score and set it as new $G$
    \end{enumerate}
    Until the score cannot be improved.
    \item Return $G$.
\end{enumerate}

\subsection{Inference}
The next task is to parametrize the learned structure by computing posterior marginal distribution of an unobserved variable $x_{v_j} \in X$ given an evidence $e=\{e_1, e_2,.. e_m\}$ on a set of variables $X(e)$. The prior marginal distribution $P(x_{v_j})$ can be computed as

\begin{equation} 
\begin{split}
P(x_{v_j}) & = \sum_{x \in X \setminus \{x_{v_j}\}} {P(X)}
\end{split}
\end{equation}
Using (1) in (3), we get
\begin{equation} \label{eq2}
\begin{split}
P(x_{v_j}) = {\sum_{x \in X \setminus \{x_{v_j}\}} \prod_{x_v \in X} P(x_v| x_{pa(v)})}
\end{split}
\end{equation}

Further evidence $e$ can be incorporated as
\begin{equation} 
\begin{split}
P(x_{v_j}|e) & = P(x_{v_j}, e) / P(e) \\
& \propto P(x_{v_j}, e) \\
& = \sum_{y \in X \setminus \{x_{v_j}\}} P(X,e) \\
& = \sum_{y \in X \setminus \{x_{v_j}\}}  \prod_{x_{v_i} \in X} P(x_{v_i}| x_{pa(v_i)}) E_e \\
& = \sum_{y \in X \setminus \{x_{v_j}\}}  \prod_{x_{v_i} \in X} P(x_{v_i}| x_{pa(v_i)}) \prod_{x \in X(e))} {E_x} \\
\end{split}
\end{equation}

For each $x_{v_j}\notin X(e)$, where $E_x$ is the evidence function for $x \in X(e)$ and $v_i$ is the node representing $x_{v_i}$. The likelihood is then defined as, 
\begin{equation}
\begin{split}
L(x_{v_j}|e) & = P(e|x_{v_j}) \\
& = \sum_{y \in X \setminus \{x_{v_j}\}}  \prod_{i \neq j} P(x_{v_i}| x_{pa(v_i)}) \prod_{x \in X(e))} {E_x} \\
\end{split}
\end{equation}
that is, the likelihood function of $x_{v_j}$ given $e$. Application of Bayes rule and using (4) and (6), then yields 

\begin{equation}
P(x_{v_j}|e) \propto L(x_{v_j}|e) P(x_{v_j}) 
\end{equation}

where the proportionality constant $P(e)$ can be computed from $P(X,e)$ by summation over $X$ and $P(x_{v_j})$ may be obtained by inference over empty set of evidence.\\

This parametrization of the learned structure with posterior marginal conditional probabilities allows to make Inference and the task is called \textit{inference-learning}. Inference learning is an extremely useful step that allow us to estimate marginal probabilities after setting evidences in the network. Depending upon the size of the network, inference-learning can be carried out with \textit{exact}, i.e. closed form or \textit{approximate} i.e. based upon Monte Carlo simulation methods. The details of these can be found in standard texts \cite{Koller2009}. In this work, we derived both exact estimates using the clique-tree algorithm and approximate estimates using \textit{rejection sampling} to estimate conditional probabilities.\vspace{\baselineskip}

\subsection{Bayesian Decision Network}
A Bayesian Network can be extended to a decision network (BDN) with the addition of \textit{utility} nodes $U$. A BDN extends a BN in the sense that a BN is a probabilistic network for belief update, where as a BDN is a probabilistic network for decision making under uncertainty. Formally defined, a BDN is a quadruple BDN:  \( N= \{ X,G,P,U \}  \) and  has three types of nodes \begin{itemize}
	\item Chance nodes ( \( X_{C} \) )\ :\ Nodes which represent events not controlled  by the decision  maker.\par

	\item Decision nodes ( \( X_{D} \) ) : Nodes which represent actions under direct control of the decision maker.\par

	\item Utility nodes (  \( X_{U}) :  \) Nodes which represent the decision maker's preference. These nodes can not be parents of chance or decision nodes.\par

\end{itemize}
\par

A decision maker interested in choosing best possible actions can either specify a structure and marginal probability distributions or use the model learned with structure and inference learning. The decision maker then ascribes a utility functions to particular states of a node. The objective of decision analysis is then to identify the decision options that maximize the expected utility. \par

The expected utility for each decision option is computed and the one which has the maximum utility output is returned. The mathematical formulation for the same can be given as follows: Let  \( A \)  be the decision variable with options  \( a_{1},a_{2}......a_{m} \) and  \( H \) is the hypothesis with states  \( h_{1},h_{2},........h_{n} \)  and  \(  \varepsilon  \) is a set of observations in the form of evidence, then the utility of an outcome (\( a_{i},h_{j}\))  is given by  \( U (a_{i},h_{j} )\) where  \( U \left( . \right)  \)  is the utility function and the expected utility is given as:\par

\vspace{\baselineskip}
\begin{equation}
EU( a_{i})  =  \sum _{j}^{}U( a_{i},h_{j}) P( h_{j} \vert  \varepsilon) _{}
\end{equation}

Where  \( P \left( . \right)  \)  is our belief in  \( H \) given  \(  \varepsilon  \) . We select the decision option  \( a^{\ast} \) which maximizes the expected utility such that:\par

\vspace{\baselineskip}
\begin{equation}
a^{\ast}= arg max_{a  \epsilon  A}EU(a)
\end{equation}

\vspace{\baselineskip}

\vspace{\baselineskip}

\section{Experiments}
\subsection{Dataset}
County-level data from \textit{The Health Inequality Project} \cite{Chetty2016}, available from \url{https://healthinequality.org/data/} were used in this study. County-level characteristics (Online Data Table 12) were merged with County-level life expectancy estimates for men and women by income quartiles (Online Data Table 11). The merged table had data on 1559 Counties and in addition to Life-expectancy estimates, it included County-level features representing (1)  \textit{Healthcare}, such as quality of preventive care, acute care, percentage of population un-insured and Medicare reimbursements (2) \textit{Health behaviors}, such as prevalence of smoking and exercise by income quartiles (3)  \textit{Income and affluence of the area}, such as median house Value and mean household income, (4)  \textit{Socioeconomic features} such as absolute upward mobility, percentage of children born to single-mothers and crime rate, (5)  \textit{Education} at the K-12 and post-secondary level, school expenditure per student, pupil-teacher ratios, test scores and income-adjusted dropout rates (6)  \textit{Demographic factors}, such as population diversity, density, absolute counts, race, ethnicity, migration, urbanization (7)  \textit{Inequality} indices, such as Gini Index, Poverty rate, Income segregation, (8)  \textit{Social cohesion indices}, such as social capital index, fraction of religious adherents in the county, (9)  \textit{Labor market conditions}, such as unemployment rate, percentage change in population since 1980, percentage change in labor force since 1980 and fraction of employed persons involved in manufacturing and (10)  \textit{Local Taxation}. Since the motivation of our work was to address the factors associated with income disparity, additional variables derived by us included (1)  \textit{Q1 - Q4 longevity-gap}, i.e., the difference in life-expectancy between income quartiles Q4 and Q1 in both males and females, (2)  \textit{Mean pooled  life-expectancy estimates} across the income quartiles Q1 through Q4 in both males and females and (3)  \textit{Pooled Standard-deviation} of life-expectancy estimates across the Q1 through Q4 income quartiles and (4)  \textit{Proportion of income quartiles Q1-Q4} in both males and females relative to the total population of the County.  Data were non-missing for most of the variables, wherever missing these were imputed with a non-parametric method for imputation in mixed-type data \cite{Stekhoven2012} implemented in  \textit{R} language for Statistical Computing \cite{RDevelopmentCoreTeam2011}. Discretization of continuous variables  was done using an in-house code written in  \textit{R} which used  \textit{k-means},  \textit{frequency-based},  \textit{quantile} and  \textit{uniform-interval} based methods in that order of preference for each variable. For each case the number of discrete classes were fixed at three for the ease of interpretation of discrete policy decisions. 
\subsection{Structure-learning and Inference}
Although the data were observational, our aim was to reason causally about some of the variables such as  \textit{State} which are known to have effects on healthcare inequities through different State policies \cite{Fisher2003,Fisher2003a,Gottlieb2010,Murray2006,Braveman2010}. In order to encode this effect in the structure of our model, we black-listed all incoming edges to State, County and Core Based Statistical Area (CBSA) prior to the start of structure-learning. The out-going edges from these nodes were allowed to be learned from the data by the structure learning algorithm. Structure learning using the hill-climbing search algorithm was repeated 1001 times on bootstrapped datasets using the \textit{R} package \textit{bnlearn} \cite{Scutari2010} and the scoring function used was \textit{Bayesian Information Criterion} (BIC). The learned 1001 structures were ensemble-averaged using majority voting criteria to arrive at the ensemble-network (Figure 1). Both exact and approximate methods were used to estimate conditional probabilities and draw inferences using the package \textit{gRain} \cite{Jsgaard2012} and \textit{bnlearn} \cite{Scutari2010} respectively. Since approximate methods rely upon MCMC sampling, we utilized this fact to repeat approximate inferences 25 times for confidence estimates using the standard deviation of estimated posterior probabilities. 

\begin{figure*}[h!]
    \centering
    \frame{\includegraphics[width=\textwidth]{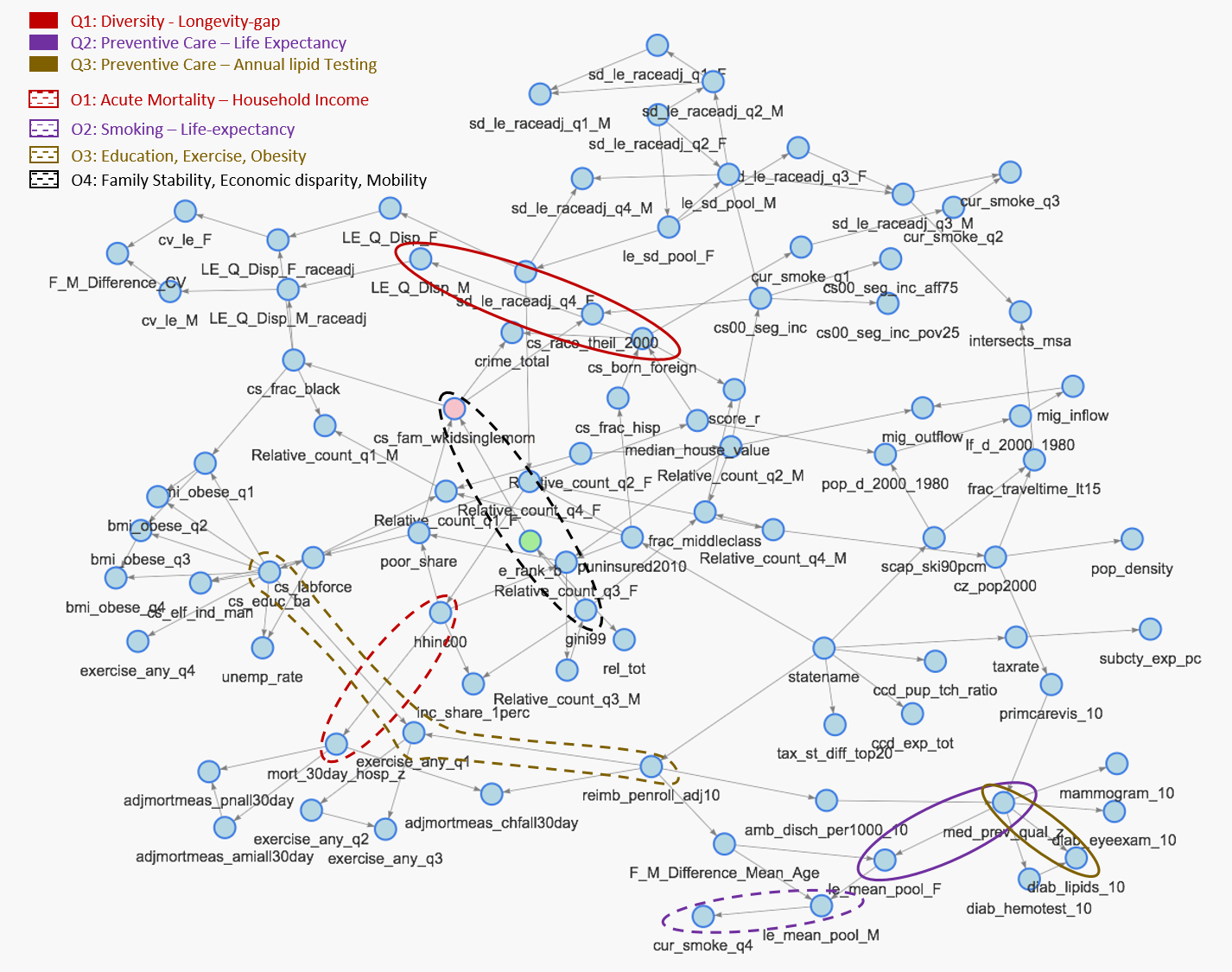}}
    \caption{\textbf{Ensemble network learned upon The Health Inequality Data}. The majority-voted structure from 1001 bootstrap structure-learning iterations used hill-climbing search along with Bayesian Information Criterion as the scoring function. Prior to learning the structures, all incoming edges into State, CBSA and County were black-listed in order to facilitate causal-reasoning about State policies and their influence upon healthcare.}
\end{figure*}

\subsection{Bayesian Decision Network}
We did not find any open source implementations that allowed a combination of data-driven structure learning with bootstraps, ensemble averaging and the extension of the learned structure to a decision network. Therefore, we wrote custom codes that allowed interfacing of the structure with available implementations that allow manual specification of a decision network \cite{Dalton2018}. We verified the consistency of network specifications and automated creation of decision networks along with their probability distributions. In order to learn optimal policy, \textit{preferences} (between -1 to +1) were defined for the states of the Utility node. For example, for assessing the factors that could minimize the longevity-gap between Q4 and Q1 income quartiles, the maximum preference (+1) was assigned to the minimum longevity-gap level (see Fig. 2, 3). Decision nodes were specified on the basis of actionable interventions (e.g. quality of preventive care in the County). Gibbs sampling \cite{Dalton2018} was used to estimate the best combination of actions (policy) that maximized the expected utility. Figure.2 shows a part of the Decision network with setting of \textit{LE\_Q\_Disp\_M} (difference in longevity between highest and lowest income quartiles in males), a derived variable defined by us.

\subsection{Interactive web-application}
We deployed the learned model as an interactive web-application developed with \textit{shinydashboard} package in \textit{R} \cite{Chang2018}. We mapped the inferences to States using the data of \textit{Global Administrative Areas} (GADM) through the use of leaflet package \cite{Cheng2018}. (\textit{Author's Note}: Screenshots of the web-application are provided in the appendix. The application will been made available on the \textit{github} page linked to this study at the time of peer-reviewed publication.)   

\begin{figure*}[h!]
    \centering
    \frame{\includegraphics[width=\textwidth]{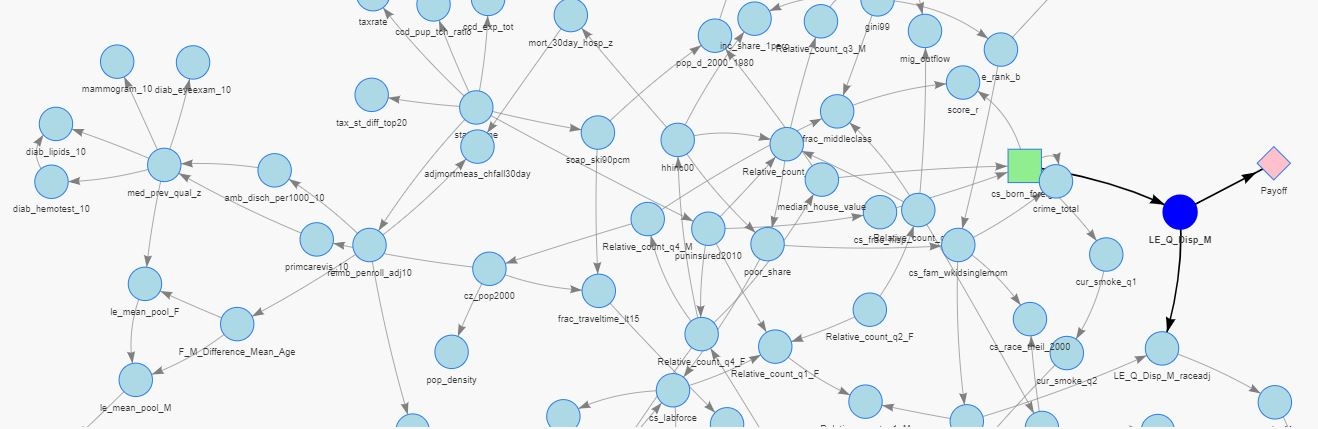}}
    \caption{\textbf{Part of Bayesian Decision Network learned from Health Inequality data}. Far right shows utilities being defined as high preference for lower disparity in life-expectancy between Q4 and Q1 income quartiles. It is seen that diversity (proportion of immigrants in the county) is a parent of this node. Policy learning demonstrated that the disparities are lower in counties with higher diversity. On the far left, the effect of preventive care services (\textit{med\_prev\_qual\_z}) on mean life expectancy in males and females is seen. Policy learning on this combination suggested that counties receiving higher percentage of Annual Lipids Test in diabetics had higher mean life expectancy both in males and females.}
\end{figure*}

\section{Results and Discussion}
With the data-driven Bayesian Network, a generative model encoding the joint probability distribution over the factors influencing longevity, we queried the network for the following \textbf{\textit{questions}}:-

\begin{figure}[h!]
    \centering
    \frame{\includegraphics[width=\columnwidth]{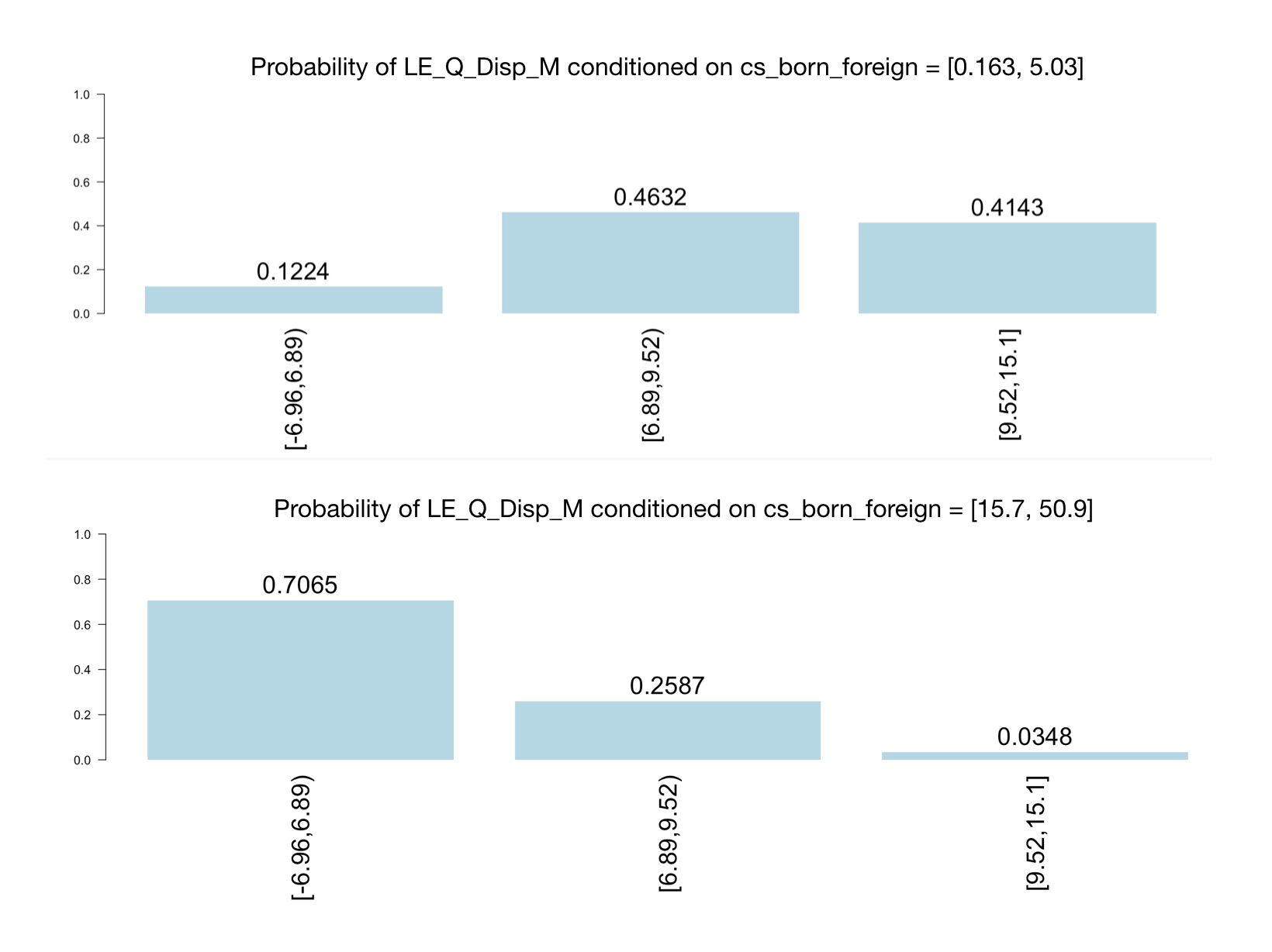}}
    \caption{\textbf{Exact inference} for estimating the effect of diversity on the longevity-gap for males. Males in Counties with lowest diversity had a 41\% probability of Q4 - Q1 longevity-gap $>$ 9 years versus males living in Counties with high diversity which showed only 3\% probability for Q4 - Q1 longevity-gap $>$ 9 years.}
\end{figure}

\noindent\textbf{(Q 1)}: What minimizes the longevity-gap between the lowest and the highest income quartiles in the Health Inequality Data?

\noindent\textbf{(A 1)}: \textit{Population diversity of the County}. We created the variables \textit{LE\_Q\_Disp\_M} and \textit{LE\_Q\_Disp\_F} as the difference in life-expectancy between income quartiles Q4 and Q1 in males and females. These represent the longevity-gap attributable to income-disparity in males and females. We observed that among all factors in the network, only \textit{cs\_born\_foreign}, the proportion of foreign-born residents in the County was a parent-node of longevity-gap. It is important to note that there might be more associations when explored through pair-wise correlations, however BNs being joint probability models, adjust for confounder, mediator and collider biases \cite{Pearl2011}. Thus the BN structure reveals that population diversity encapsulates all the factors that may be indirect influences upon longevity-gap. Exact inference with setting evidence on different levels of diversity revealed that males in Counties with lowest diversity had a 41\% probability of Q4 - Q1 longevity-gap $>$ 9 years versus males living in Counties with high diversity which showed only 3\% probability for Q4 - Q1 longevity-gap $>$ 9 years. The model structure also revealed that high diversity was a first-degree neighbor of higher \textit{median house value} and second degree neighbor of \textit{proportion of high-earning females} in the County. Interestingly, Chetty et al observed that beyond a certain threshold, increasing income does not yield proportionate gains in longevity, and our inference may be indicative of the same effect. Thus our model captures this observation along with more nuances in the network and is able to provide quantitative estimates for these effects as inferred from the joint probability model.

\noindent\textbf{(Q 2)}: What maximizes the mean life expectancy in males and females?

\noindent\textbf{(A 2)}: \textit{Preventive Care Quality}. We observed that \textit{med\_prev\_qual\_z}, the \textit{Index of Preventive Care}, was the only parent-node of \textit{le\_mean\_pool\_F} and a grand-parent of \textit{le\_mean\_pool\_M} (variables derived by us from the Health Inequality Data). Estimates drawn through exact inference reveal that high quality Preventive Care \textit{med\_prev\_qual\_z} improves the probability of living beyond 85 years of age by a staggering 43\% in females and 30\% in males. Preventive Quality Indices (PQI) provide a proxy for healthcare quality of the system outside the hospital setting and were compiled from the Dartmouth Atlas as a part of The Health Inequality Project. PQIs are based upon "ambulatory care sensitive conditions" (ACSCs) such as diabetes, i.e., conditions in which high-quality outpatient care or early interventions can prevent hospitalizations and complications. PQIs are used along with \textit{discharges for ASCS per thousand} and the association between these was captured as a first-order relation between these variables in our graphical model. Therefore, improving PQIs is the most actionable step for increasing mean life-expectancy and for reducing economic burden due to hospitalizations as indicated by our model.

\begin{figure}[h!]
    \centering
    \frame{\includegraphics[width=\columnwidth]{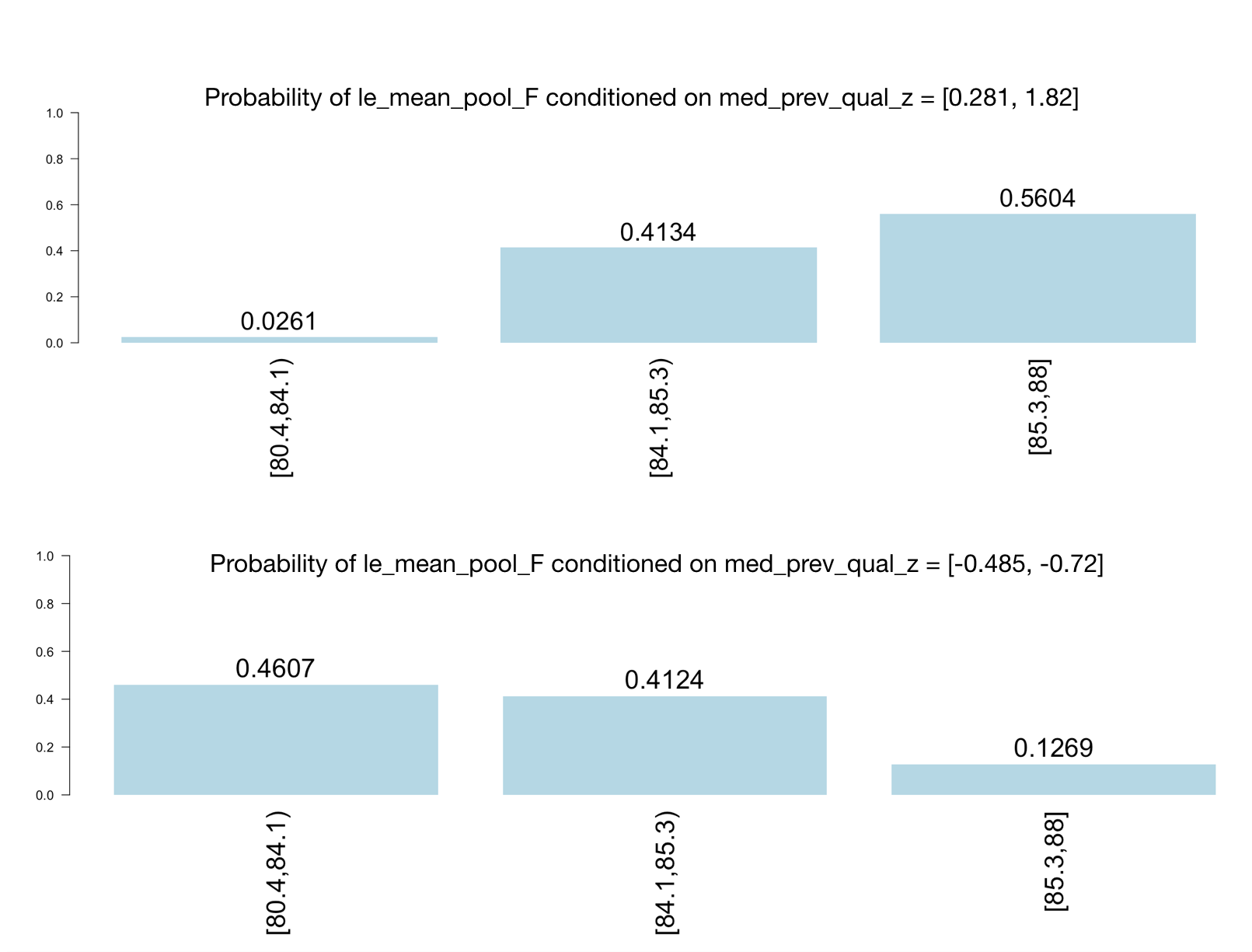}}
    \caption{\textbf{Exact-inference} based estimates for influence of preventive care on pooled life expectancy in females. A difference of 43.3\% (0.56 - 0.13) was seen in probability of life-expectancy beyond 85 years of age. Similar analysis in males revealed a difference of 30\%.}
\end{figure}

\noindent\textbf{(Q 3)}: Which Preventive Care measure maximizes the probability of life-expectancy beyond 85 years? 

\noindent\textbf{(A 3)}: \textit{Annual Lipid Testing in the diabetic population}. We asked this question as a policy-learning question from the perspective of maximizing the availability of these tests during the preventive care visits. This was pertinent as Medicare reimbursements were found to be drastically different across the states (visualized as a heatmap on the web-application) which in-turn were linked to the quality of preventive care. The data included PQI indicators only for diabetes and mammography. We set a high preference on high longevity as the utility node and PQIs as decision nodes. The policy table learned from simulations (Table 1) indicates that the payoff was maximized by focusing on Annual Lipid Testing in the proportion of population that was diabetic.

\begin{table}[t]
\resizebox{\columnwidth}{!}
{\begin{tabular}{|l|l|l|r|}
\hline
\textbf{diab\_eyeexam\_10}    & \textbf{mammogram\_10}    & \textbf{diab\_lipids\_10} & \textbf{payoff} \\
\hline
[70.2,85.6]    & [68.2,86.1]    &[79.3,92.9] & 0.52\\
\hline
[70.2,85.6]    & [68.2,86.1]    &[65.6,79.3) & 0.48\\
\hline
[62.2,70.2)    & [68.2,86.1]    &[79.3,92.9] & 0.44\\
\hline
[70.2,85.6]    & [59.5,68.2)    &[79.3,92.9] & 0.40\\
\hline
[62.2,70.2)    & [68.2,86.1]    &[65.6,79.3) & 0.26\\
\hline
[70.2,85.6]    & [59.5,68.2)    &[65.6,79.3) & 0.22\\
\hline
[42.4,62.2)    & [68.2,86.1]    &[79.3,92.9] & 0.20\\
\hline
[62.2,70.2)    & [59.5,68.2)    &[79.3,92.9] & 0.12\\
\hline
[70.2,85.6]    & [31.1,59.5)    &[79.3,92.9] & 0.11\\
\hline
[42.4,62.2)    & [68.2,86.1]    &[65.6,79.3) & 0.08\\
\hline
\end{tabular}}

    \caption{\label{Table 2}\textbf{Policy table} learned by setting maximum preference on longevity beyond 85 years in females. The table indicates that keeping Annual Lipid Testing in the highest stratum is expected to maximize this objective. We infer this from the absence of lowest stratum in this test among the top 10 policy combinations ranked by payoffs}
\end{table}

In addition to the directed questions, the graphical decision model allowed us to make the following \textbf{\textit{observations}}:-

\noindent\textbf{(O 1)}: \textit{Acute mortality and mean household-income}. We observed that mean household income is a direct (and only) parent of 30-day hospital mortality index in our model. Areas with high mean household-income (greater than \$45,000 p.a.) have a 30\% less probability of having high 30-day Hospital Mortality Rate Index (greater than 0.92) as compared to areas with low mean household-income (less than \$30,000 p.a.). Tracing the grand-children nodes of Hospital Mortality Rate Index, \textit{Pneumonia} had the highest contribution to this effect among other available diseases including congestive heart failure and acute myocardial infarction.

\noindent\textbf{(O 2)}: \textit{Smoking and mean life-expectancy}. We observed that smoking was a child-node of mean life expectancy in males in the network, i.e. Counties with lower proportions of currently smoking males in Q4 income quartile showed a 30\% increase in probability of living beyond 82 years.

\noindent\textbf{(O 3)}: \textit{Education, Exercise, Obesity and Longevity}. We observed that graduate level education \textit{cs\_educ\_ba} was a major distributor of probabilistic influence in the network and was linked to obesity, exercise, income and unemployment rates. This indicates that a significant part of the effect of exercise and obesity can be apportioned to education as the driver of healthy behaviors and higher income. Although these findings are not surprising, our model reveals these in a transparent, unified manner and allows inference queries to estimate quantitative effects of these factors on health outcomes. For example, we estimated that areas with exercising populations, especially in Q1 income quartile, have a 19\% lower probability of hospitalization rates in the highest band after correcting influences from other variables present in the data.

\noindent\textbf{(O 4)}: \textit{Poverty breeds poverty, links with racial factors and stability of families}. In addition to health-inequities, our model illuminated the social disparities and their indirect role in propagating health inequity through income disparities. We observed that high income-segregation (\textit{Gini index}) was a parent of and negatively associated with \textit{absolute upward mobility} i.e. the upward mobility in percentage of children born to lower quartile income parents. This indicates that poverty was associated with lower inter-generational mobility, thus perpetuating the socio-economic disparities in the society which are well studied in the United States \cite{Levy1989}. Our model also confirmed the univariate correlations between low social mobility being linked with lower family stability(40\% lowered mobility) and higher Gini disparity (37\% lowered mobility) as indicated by \cite{Chetty2016}. The latter phenomenon referred to as \textit{assortative mating} or the "marriage-gap"  has been noticed to consistently increase in the recent years in the United States and is an under-appreciated factor in widening income and health disparities.

\section{Conclusion}
This study presents the overarching potential for using data-driven graphical models extended to decision networks as a unified framework for enabling healthcare policy. We demonstrate this potential through the use-case of developing a coherent understanding of health-inequality in the United States given from a heterogenous dataset. Our proposed framework for understanding the longevity-gap uses a generative modeling approach to gain a system-level understanding of healthcare inequities. It also emphasizes transparency and explainability and extends this motivation through creation of a web-application that encapsulates our model inferences and visualizations for general users and policymakers. We expect that the users will not only be able to validate our findings but also explore further insights into healthcare and social inequity that we may have missed. In conclusion, with this study we reason that Artificial Intelligence research has the potential to reduce disparities and recommend actionable solutions for promote individual and social health.

\section{Acknowledgements}
We acknowledge the inputs and support provided by Dr. Nigam Shah, Biomedical Informatics Research, Stanford University, USA and Dr. Rakesh Lodha, Department of Pediatrics, All India Institute of Medical Sciences, New Delhi, India. This work was supported in part by the Wellcome Trust/DBT India Alliance Early Career Award IA/CPHE/14/1/501504 to Tavpritesh Sethi.

\bibliographystyle{aaai}\bibliography{Main_Paper.bib}

\begin{thebibliography}{}

\bibitem[\protect\citeauthoryear{Abr{\`{a}}moff \bgroup et al\mbox.\egroup
  }{2018}]{Abramoff2018}
Abr{\`{a}}moff, M.~D.; Lavin, P.~T.; Birch, M.; Shah, N.; and Folk, J.~C.
\newblock 2018.
\newblock {Pivotal trial of an autonomous AI-based diagnostic system for
  detection of diabetic retinopathy in primary care offices}.
\newblock {\em npj Digital Medicine} 1(1):39.

\bibitem[\protect\citeauthoryear{AS and Smith}{2018}]{AS2018}
AS, A., and Smith, A.
\newblock 2018.
\newblock {Machine learning and health care disparities in dermatology}.
\newblock {\em JAMA Dermatology}.

\bibitem[\protect\citeauthoryear{Braveman \bgroup et al\mbox.\egroup
  }{2010}]{Braveman2010}
Braveman, P.~A.; Cubbin, C.; Egerter, S.; Williams, D.~R.; and Pamuk, E.
\newblock 2010.
\newblock {Socioeconomic disparities in health in the united States: What the
  patterns tell us}.
\newblock {\em American Journal of Public Health}.

\bibitem[\protect\citeauthoryear{Cabitza, Rasoini, and
  Gensini}{2017}]{Cabitza2017}
Cabitza, F.; Rasoini, R.; and Gensini, G.~F.
\newblock 2017.
\newblock {Unintended Consequences of Machine Learning in Medicine}.
\newblock {\em JAMA}.

\bibitem[\protect\citeauthoryear{Chang and {Borges Ribeiro}}{2018}]{Chang2018}
Chang, W., and {Borges Ribeiro}, B.
\newblock 2018.
\newblock {\em {shinydashboard: Create Dashboards with 'Shiny'}}.

\bibitem[\protect\citeauthoryear{Cheng, Karambelkar, and Xie}{2018}]{Cheng2018}
Cheng, J.; Karambelkar, B.; and Xie, Y.
\newblock 2018.
\newblock {\em {leaflet: Create Interactive Web Maps with the JavaScript
  'Leaflet' Library}}.

\bibitem[\protect\citeauthoryear{Chetty \bgroup et al\mbox.\egroup
  }{2016}]{Chetty2016}
Chetty, R.; Stepner, M.; Abraham, S.; Lin, S.; Scuderi, B.; Turner, N.;
  Bergeron, A.; and Cutler, D.
\newblock 2016.
\newblock {The association between income and life expectancy in the United
  States, 2001-2014}.
\newblock {\em JAMA - Journal of the American Medical Association}.

\bibitem[\protect\citeauthoryear{Dalton and Nutter}{2018}]{Dalton2018}
Dalton, J.~E., and Nutter, B.
\newblock 2018.
\newblock {\em {HydeNet: Hybrid Bayesian Networks Using R and JAGS}}.

\bibitem[\protect\citeauthoryear{Fisher \bgroup et al\mbox.\egroup
  }{2003a}]{Fisher2003}
Fisher, E.~S.; Wennberg, D.~E.; Stukel, T.~A.; Gottlieb, D.~J.; Lucas, F.~L.;
  and Pinder, {\'{E}}.~L.
\newblock 2003a.
\newblock {The implications of regional variations in Medicare spending. Part
  1: The content, quality, and accessibility of care}.
\newblock {\em Annals of Internal Medicine}.

\bibitem[\protect\citeauthoryear{Fisher \bgroup et al\mbox.\egroup
  }{2003b}]{Fisher2003a}
Fisher, E.~S.; Wennberg, D.~E.; Stukel, T.~A.; Gottlieb, D.~J.; Lucas, F.~L.;
  and Pinder, {\'{E}}.~L.
\newblock 2003b.
\newblock {The implications of regional variations in Medicare spending. Part
  2: Health outcomes and satisfaction with care}.
\newblock {\em Annals of Internal Medicine}.

\bibitem[\protect\citeauthoryear{Geiger, Verma, and Pearl}{1990}]{Geiger1990}
Geiger, D.; Verma, T.; and Pearl, J.
\newblock 1990.
\newblock {Identifying independence in bayesian networks}.
\newblock {\em Networks}.

\bibitem[\protect\citeauthoryear{Gottlieb \bgroup et al\mbox.\egroup
  }{2010}]{Gottlieb2010}
Gottlieb, D.~J.; Zhou, W.; Song, Y.; Andrews, K.~G.; Skinner, J.~S.; and
  Sutherland, J.~M.
\newblock 2010.
\newblock {Prices don't drive regional Medicare spending variations}.
\newblock {\em Health Affairs}.

\bibitem[\protect\citeauthoryear{Jsgaard}{2012}]{Jsgaard2012}
Jsgaard, S.~H.
\newblock 2012.
\newblock {Graphical Independence Networks with the gRain Package for R}.
\newblock {\em Journal of Statistical Software}.

\bibitem[\protect\citeauthoryear{Keane and Topol}{2018}]{Keane2018}
Keane, P.~A., and Topol, E.~J.
\newblock 2018.
\newblock {With an eye to AI and autonomous diagnosis}.
\newblock {\em npj Digital Medicine} 1(1):40.

\bibitem[\protect\citeauthoryear{Koller and Friedman}{2009}]{Koller2009}
Koller, D., and Friedman, N.
\newblock 2009.
\newblock {\em {Probabilistic Graphical Models: Principles and Techniques}},
  volume 2009.

\bibitem[\protect\citeauthoryear{LaVeist, Gaskin, and
  Richard}{2011}]{LaVeist2011}
LaVeist, T.~A.; Gaskin, D.; and Richard, P.
\newblock 2011.
\newblock {Estimating the Economic Burden of Racial Health Inequalities in the
  United States}.
\newblock {\em International Journal of Health Services}.

\bibitem[\protect\citeauthoryear{Levy and Wilson}{1989}]{Levy1989}
Levy, F., and Wilson, W.~J.
\newblock 1989.
\newblock {The Truly Disadvantaged}.
\newblock {\em Journal of Policy Analysis and Management}.

\bibitem[\protect\citeauthoryear{Murray \bgroup et al\mbox.\egroup
  }{2006}]{Murray2006}
Murray, C.~J.; Kulkarni, S.~C.; Michaud, C.; Tomijima, N.; Bulzacchelli, M.~T.;
  Iandiorio, T.~J.; and Ezzati, M.
\newblock 2006.
\newblock {Eight Americas: Investigating mortality disparities across races,
  counties, and race-counties in the United States}.
\newblock {\em PLoS Medicine}.

\bibitem[\protect\citeauthoryear{Pearl}{1985}]{Pearl1985a}
Pearl, J.
\newblock 1985.
\newblock {Bayesian Networks A Model of Self-Activated Memory for Evidential
  Reasoning}.

\bibitem[\protect\citeauthoryear{Pearl}{2011}]{Pearl2011}
Pearl, J.
\newblock 2011.
\newblock {\em {Causality: Models, reasoning, and inference, second edition}}.

\bibitem[\protect\citeauthoryear{{R Development Core
  Team}}{2011}]{RDevelopmentCoreTeam2011}
{R Development Core Team}, R.
\newblock 2011.
\newblock {\em {R: A Language and Environment for Statistical Computing}},
  volume~1.

\bibitem[\protect\citeauthoryear{Scutari}{2010}]{Scutari2010}
Scutari, M.
\newblock 2010.
\newblock {Learning Bayesian Networks with the bnlearn R Package}.
\newblock {\em Journal of Statistical Software} 35(3):1--22.

\bibitem[\protect\citeauthoryear{Stekhoven and
  B{\"{u}}hlmann}{2012}]{Stekhoven2012}
Stekhoven, D.~J., and B{\"{u}}hlmann, P.
\newblock 2012.
\newblock {Missforest-Non-parametric missing value imputation for mixed-type
  data}.
\newblock {\em Bioinformatics}.

\end{thebibliography}

\section{Supplementary File (submitted separately)}

\subsection{\textbf{Screenshots of the web-application}}
\subsubsection{Structure.}

The web-application provides interactive visualizations including leaflet maps for geographic regions of the United States that can be explored for interactive inference. The dashboard will be made available as a web-application along with the paper. The Graphical User Interface of the dashboard is shown in Figure 5.

\begin{figure*}[h!]
    \centering
    \frame{\includegraphics[width=\textwidth]{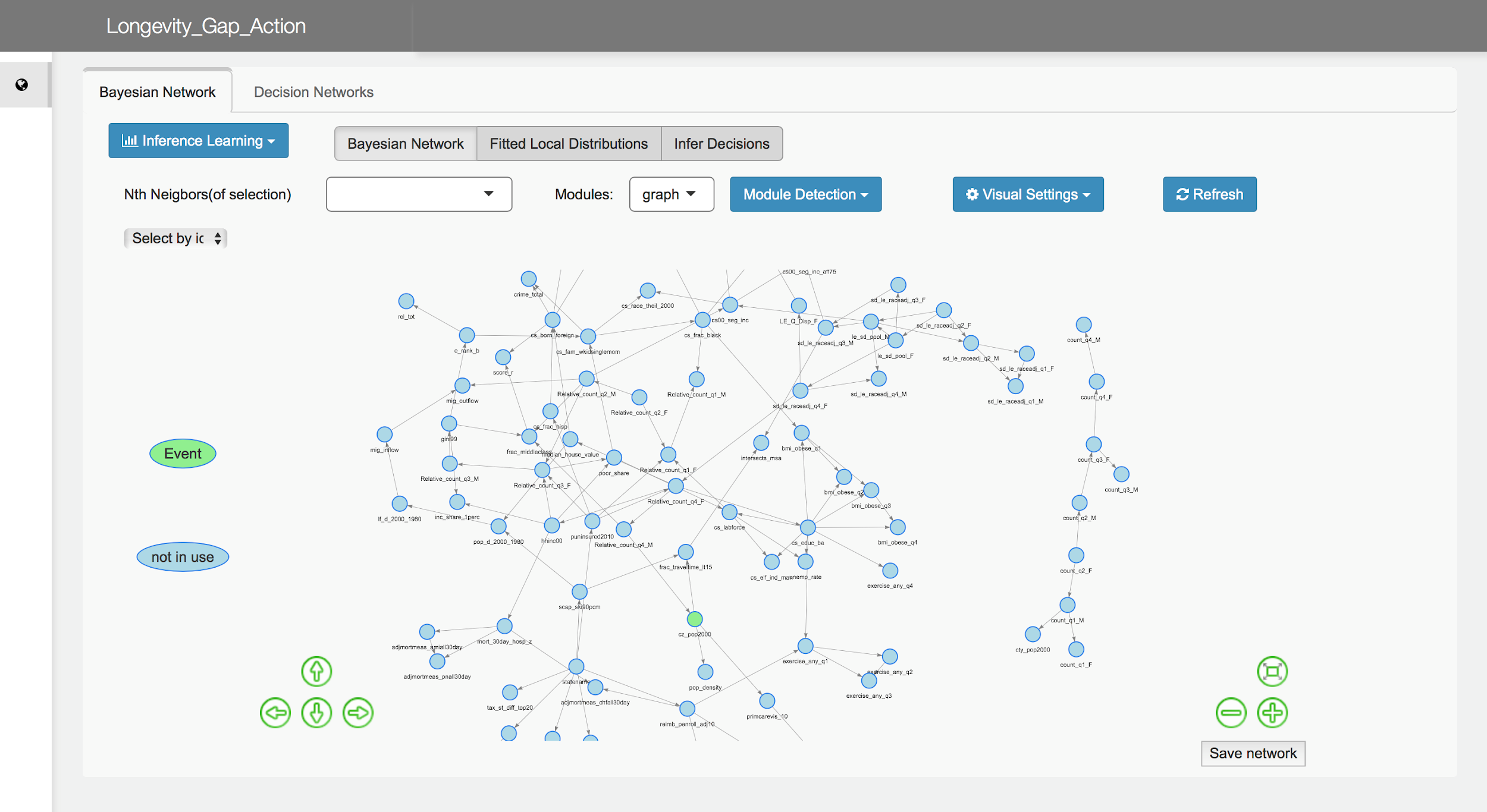}}
    \caption{\textbf{Screenshot of the dashboard.}}
\end{figure*}

\subsubsection{\textbf{Inference.}}
The inference learning menu, shown in Figure 6, enables the user to learn and explore conditional probability plots on the learned structure. The user can set an event and multiple evidences can be inserted and removed. This crucial feature enables the user to explore probability plots on event nodes in the network conditionalized on number  of evidences.
The web-application allows two ways to perform for inference learning, 
\begin{itemize}
    \item Approximate Inference. Faster to learn on large networks, but time adds up for each inference as it relies upon Monte Carlo simulations for each. We add the option of repeating the sampling 25 times to get an estimate of error bars for each inference.
    \item Exact inference for smaller networks. One-time learning, may be intractable for large networks but is faster when inferences need to be repeatedly assessed as it avoids Monte Carlo sampling every time.
\end{itemize}

While the web-application learns approximate learning as default, user must explicitly learn exact inferences whenever the structure is learned or updated. We report exact inferences in this paper. 

\begin{figure*}[h!]
    \centering
    \frame{\includegraphics[width=\textwidth]{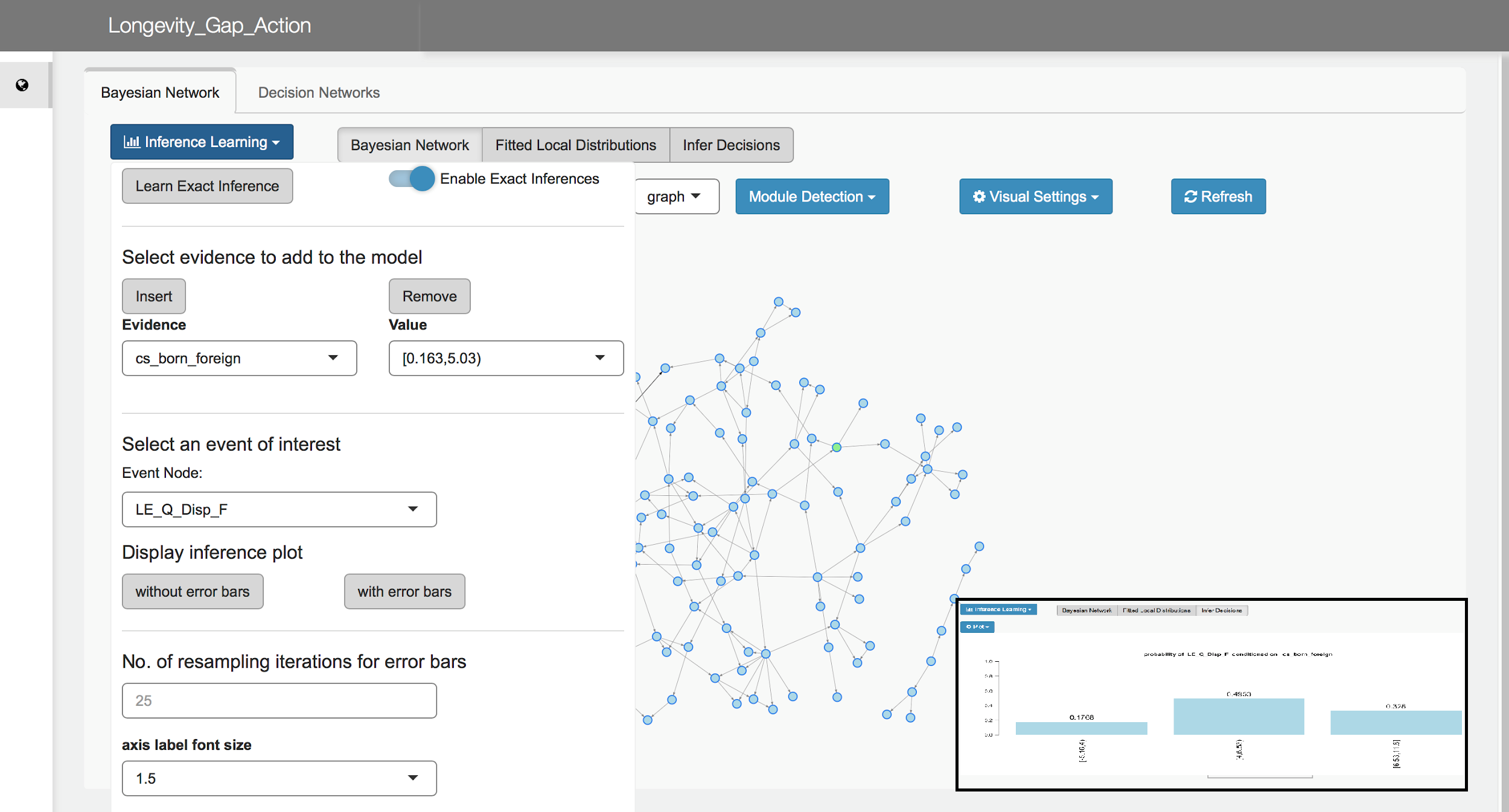}}
    \caption{\textbf{Setting Inferences. We learned Exact Inference for the work presented in this paper.The inset shows an example of a learned inference.}}
\end{figure*}

\subsubsection{\textbf{Bayesian Decision Network.}}
Figure 7 shows the Decision Network constructed by setting the decision nodes for optimal policy learning.

\begin{figure*}[h!]
    \centering
    \frame{\includegraphics[width=\textwidth]{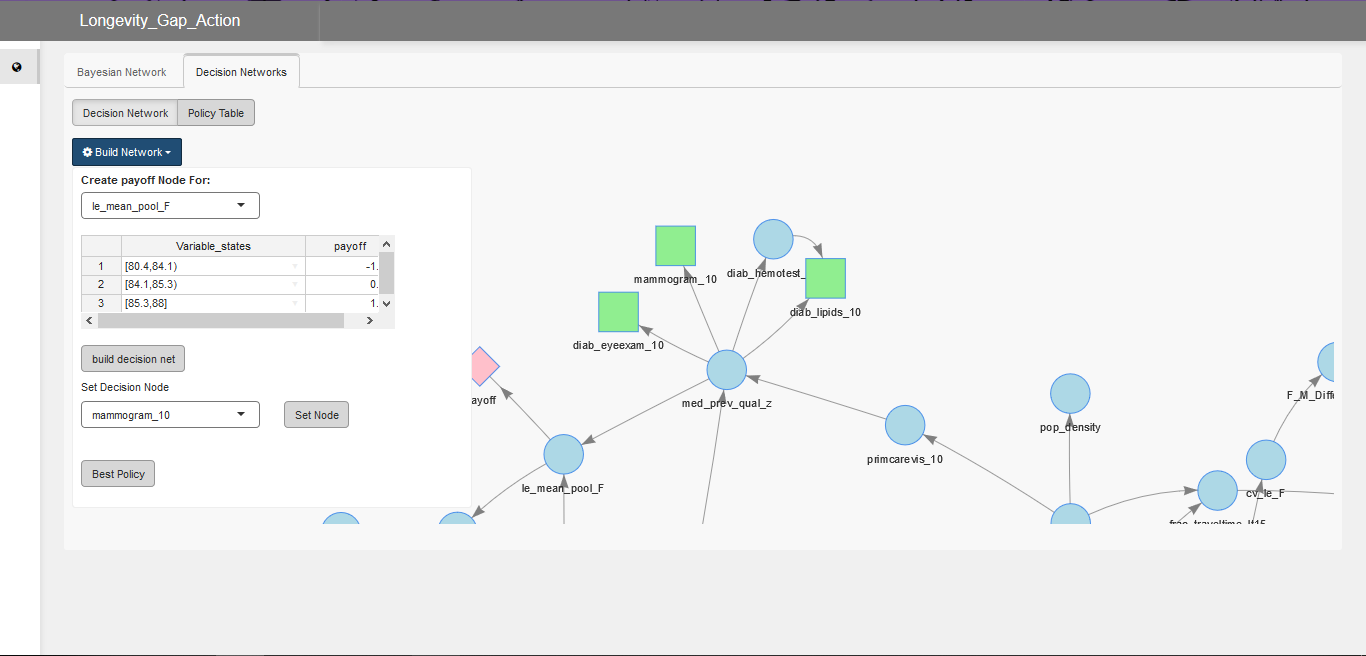}}
    \caption{\textbf{Setting the utility and specification of decision nodes for Decision Network.}}

\end{figure*}

\subsubsection{\textbf{Inferring Policy Actions.}}
Figure 8 shows an example of the learned policy on the web-application.

\begin{figure*}[h!]
    \centering
    \frame{\includegraphics[width=\textwidth]{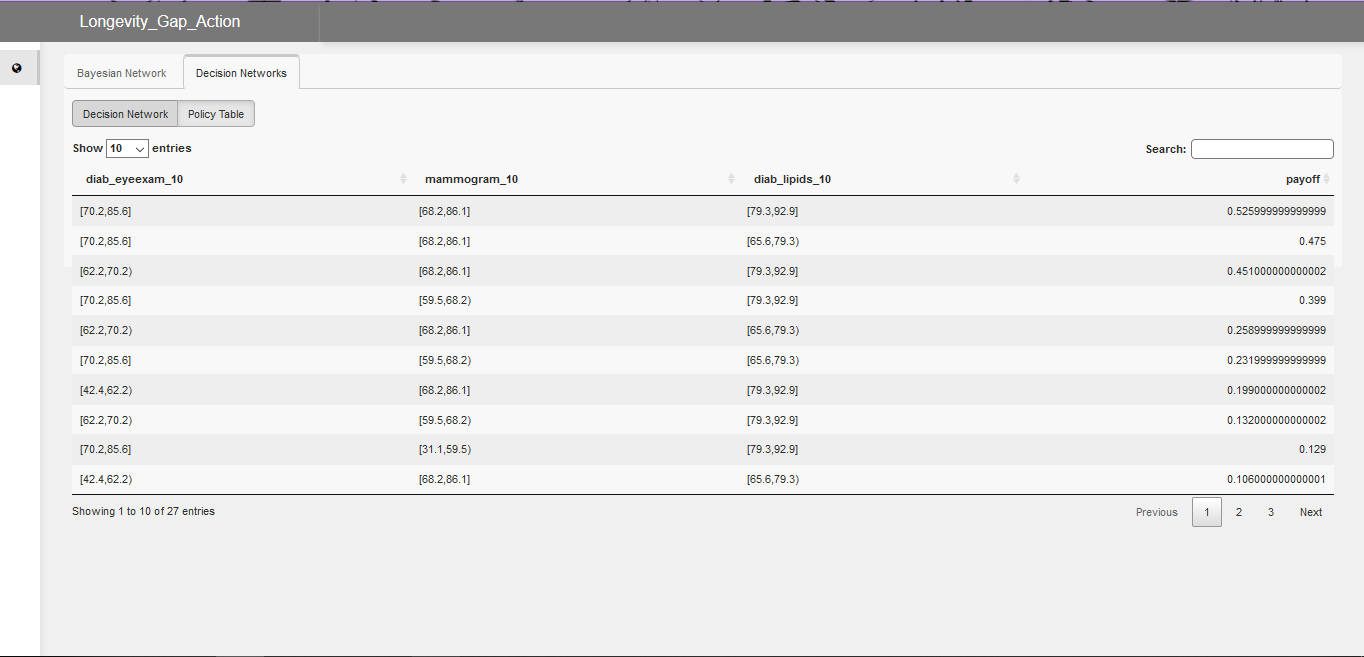}}
    \caption{\textbf{Example of a learned policy table on the web-application.}}

\end{figure*}

\end{document}